\documentclass[11pt]{JHEP3}

\skip\footins = 1\bigskipamount plus 2pt minus 4pt

\def\ind{\mathbb{E}}
\def\sig{\mathbb{S}}
\def\d{{\partial}}

\newcommand{\cN}{{\cal N}}
\newcommand{\cF}{{\cal F}}
\newcommand{\cC}{{\cal C}}

\newcommand{\cH}{{\cal H}}
\newcommand{\hZ}{\mathbb{Z}}
\newcommand{\rH}{{\textrm{H}}}
\newcommand{\tr}{\mathop{\rm Tr}\nolimits}
\newcommand{\di}{\mathop{\rm dim}\nolimits}
\newcommand{\rim}{\mathop{\rm Im}\nolimits}
\newcommand{\rker}{\mathop{\rm Ker}\nolimits}
\newcommand{\lp}{\left(}
\newcommand{\rp}{\right)}

\def\inc#1#2{\langle #1, #2 \rangle_{\cC}}
\def\inn#1#2{\langle #1, #2 \rangle}

\def\cU#1{{\cal U_{\mathit{#1}}}}
\def\cD#1{{\cal D}_{\mathit{#1}}}
\def\cCo#1{{\cal C}_{\mathit{#1}}}

\newcommand{\SL}{\mathop{\rm SL}\nolimits}
\newcommand{\U}{\mathop{\rm {}U}\nolimits}
\newcommand{\SU}{\mathop{\rm SU}\nolimits}
\newcommand{\R}{\mathbb{R}}
\newcommand{\trk}{{\tr_{\cN}\left(q^{L_0}\right)}}

\title{BRST quantization of string theory in $AdS_3$}

\author{Ari Pakman\\
Racah Institute of Physics, The Hebrew University\\
Jerusalem 91904, Israel\\
E-mail: \email{pakman@phys.huji.ac.il}}

\abstract{We study the BRST quantization of bosonic and NSR strings
propagating in $AdS_3 \times {\cal N}$ backgrounds. The no-ghost
theorem is proved using the Frenkel-Garland-Zuckerman method. Regular
and spectrally-flowed representations of affine ${\rm
SL}(2,\mathbb{R})$ appear on an equal footing. Possible
generalizations to related curved backgrounds are discussed.}

\begin{document}

\section{Introduction}\label{section1}

The BRST quantization of gauge theories was first applied to string
theory in~\cite{Fujikawa:1981yy,HH1,Ohta:af}, and in~\cite{Kato:1982im} a
proof of the no-ghost theorem was given in the BRST context. Once the
physical space was identified with the cohomology of the BRST operator
at a particular ghost number, the cohomology itself is built more or
less explicitly, in order to show that it does not contain states of
negative norm. In particular, in~\cite{Kato:1982im} it was shown that,
for strings in flat space, physical states can be isometrically
identified with the excitations in the light-cone gauge.

In~\cite{FGZ} Frenkel, Garland and Zuckerman (FGZ) presented a
technically simple but conceptually deep new proof of the no-ghost
theorem in the BRST context, which relies heavily on the fact that the
\emph{relative} BRST cohomology\footnote{The relative BRST cohomology
is defined imposing the additional constraints $b_0=L_0=0$ to the BRST
cohomolgy (for fermionic strings in the Ramond sector, also
$\beta_0=G_0=0$ ). See appendix~\ref{section6}.}  at non-zero momentum
is concentrated at one ghost number,\footnote{As shown in~\cite{FGZ}
(and in~\cite{Figueroa-O'Farrill:1988hu} for the NSR string), the
cohomology of the \emph{full} BRST complex at nonzero momentum
consists of two isomorphic copies at consecutive ghost numbers.} a
result known as the ``vanishing theorem''.  Once the latter
holds,\footnote{It should be noted that the vanishing theorem does not
hold in less conventional settings.  For example, the BRST analysis of
2D gravity coupled to $c \leq 1$ matter shows that physical states are
distributed through several ghost numbers~\cite{Bouwknegt:1991yg}.}
the proof of the no-ghost theorem is reduced to showing the equality
between the traces of two operators (the method was discussed
in~\cite{Spiegelglas}).  The power of the FGZ method lies on its
economy: it probes the unitarity of the physical sector without
explicitly building it.  The method was applied to bosonic strings
in~\cite{FGZ,Spiegelglas}, to NRS strings
in~\cite{Figueroa-O'Farrill:1988hu}, to ${\R}^{1,d-1} \times \cN$ ($d
\geq 2$) backgrounds in~\cite{Zuckerman:1987ds} and to $\R^{1} \times
\cN$ backgrounds in~\cite{Asano}.

The FGZ proof of the vanishing theorem~\cite{FGZ} and its NSR
extensions in~\cite{Figueroa-O'Farrill:1988hu}, use techniques of
homological algebra and do not depend on the details of the chiral
algebra of the matter conformal field theory (CFT).  The only
assumption needed is that the matter CFT should be an $L_-$-free
Virasoro module.\footnote{This means that the Hilbert space of the
chiral CFT can be expressed as a direct sum of spaces, each of which
is built by the action of Virasoro $L_{n <0}$ operators on a Virasoro
highest-weight state, such that states with different $L_{n <0}$
content are linearly independent.  In the superconformal case, the
operators $G_{n<0}$ should be included.  The importance of this
condition can be seen in a hand-waving way as follows.  Since the BRST
charge $Q$ is built out of the Virasoro modes only (and knows nothing
about the worldsheet chiral algebra), certain homological statements
about $Q$ (such as the vanishing theorem), can be proved when a basis
for the space upon which $Q$ acts, can itself be obtained from
Virasoro modes only.}  For flat space worldsheet CFT, the latter was
shown in~\cite{Brower}.

In this work we will apply the FGZ method to bosonic and NSR strings
propagating in $AdS_3 \times \cN$
backgrounds~\cite{Malda,Malda2}. Here $\cN$ is a compact unitary CFT,
with the necessary central charge for the worldsheet CFT to be
critical.  These vacua are relevant in the context of the $AdS/CFT$
correspondence~\cite{Aharony:1999ti}.  In the $AdS_3/CFT_2$ case, the
infinite modes of the two-dimensional (super)conformal algebras in the
dual boundary CFT can be constructed explicitly
(see~\cite{GKS,Giveon:1999jg,Argurio:2000tb,Giveon:2003ku} and
references therein).

The vanishing theorem holds in $AdS_3 \times \cN$ vacua, because the
worldsheet CFT is an $L_-$-free Virasoro module, as shown
in~\cite{Hwang:1991an,Evans:1998qu} for regular $\SL(2,\R)$
representations and in~\cite{Malda,Pakman:2003cu} for the spectrally
flowed cases.\footnote{For fermionic strings,
refs.~\cite{Evans:1998qu,Pakman:2003cu} consider only the NS
sector. The results can be easily extended to the R sector.
Alternatively, one can consider the $AdS_3 \times \cN$ backgrounds as
a particular case of the backgrounds studied in~\cite{Asano}, the
timelike $\U(1)$ current being the $J^3(z)$ current of $\SL(2,\R)$.
But since the whole $AdS_3 \times \cN$ CFT is an $L_-$-free Virasoro
module, the new filtration introduced in~\cite{Asano} (which considers
only the timelike $\U(1)$, $c=1$ CFT), is not strictly necessary.}

As in the $\R^{1} \times \cal{N}$ vacua studied in~\cite{Asano},
$AdS_3 \times \cN$ is a nontrivial setting to see the BRST
quantization at work since, due to the absence of a lightcone
direction in $AdS_3$, we cannot identify the action of the ghost
degrees of freedom as annihilating the ``transversal'' string
excitations.

The unitarity puzzles posed by string theory in $AdS_3$
backgrounds~\cite{Balog:1988jb} were settled in the old covariant
quantization (OCQ) formalism
in~\cite{Hwang:1991an,Henningson:1990ua,Evans:1998qu,Pakman:2003cu}. It
was shown there that physical states belong to the $\SL(2,\R) /\U(1)$
coset, modulo spurious states. This reduces the problem to the
unitarity of the coset representations, which was proved in~\cite{DPL}
and~\cite{Pakman:2003cu} for the bosonic and fermionic cases,
respectively. Coset unitarity requires a bound for the the spin of the
discrete representations of $\SL(2,\R)$. This bound is rather
ubiquitous, and we discuss it in section~\ref{section3.1}.  As we show
in section~\ref{section4}, one important point in which the OCQ
formalism differs from the FGZ approach, is that for fermionic NRS
strings, in the FGZ proof there is no need for the unitarity of the
\emph{supersymmetric} $\SL(2,\R) /\U(1)$ coset~\cite{Pakman:2003cu},
the bosonic results of~\cite{DPL} being sufficient.  A different
approach to string unitarity in $AdS_3$ has been advocated
in~\cite{Bars:1995mf}.

Both in the OCQ and FGZ approaches, the general results do not hold
when $J^3_0=0$, since the matter CFT is no longer an $L_-$-free
module.  The unitarity of physical states must then be checked by
hand.\footnote{This is similar to the zero-momentum states in flat
space.}  These exceptional cases have been already studied
in~\cite{Malda,Evans:1998qu,Pakman:2003cu} for different
representations, and will not be considered here.

The plan of this work is as follows.  In section~\ref{section2} we
review the string spectrum in $AdS_3$.  In section~\ref{section3} we
review the Hodge theory of the string BRST complex, which is an
elegant and clear setting in which to formulate the FGZ method, which
we do in section~\ref{section3.1}.  In section~\ref{section4} we
calculate all the traces needed to prove the no-ghost theorem for the
different representations.  Finally, in section~\ref{section5} we
discuss possible generalizations to other curved backgrounds.  In
appendix~\ref{section6} we summarize the essentials of the BRST
formalism needed in this work.  Appendices~\ref{section7}
and~\ref{section8} are devoted to technical points used in the body of
the paper. We work in the holomorphic sector.  For closed strings, a
similar antiholomorphic copy should be considered.

\section{The affine $\SL(2,\R)$ theory}\label{section2}

\subsection{The bosonic algebra}\label{section2.1}

The space $AdS_3$ is the universal cover of the group manifold
$\SL(2,\R)$, so the action of a string moving in an $AdS_3$ background
is an $\SL(2,\R)$ WZW model. The symmetry algebra is the affine
$\SL(2,\R)$ algebra at level $k$ ($k>2$), generated by three currents
$J^{3,\pm}(z)$ satisfying the OPEs
\begin{eqnarray}
J^3(z) J^3(w) &\sim &  -{k/2 \over (z-w)^2} \,,
\nonumber\\
J^+(z) J^-(w) &\sim &  {k \over (z-w)^2} + {2 J^3(w) \over z-w}\,,
\nonumber\\
J^3(z) J^{\pm}(w) &\sim &  \pm { J^{\pm}(w) \over z-w}\,.
\label{opeads}
\end{eqnarray}
Using the Sugawara construction, the stress tensor is
\begin{equation}
\label{sugawara}
T^{{\SL(2,\R)}} = {1\over k-2} \frac12 \left(J^+J^- + J^-J^+
-2J^3J^3 \right) .
\end{equation}
and its modes close a Virasoro algebra with central charge
$c_{\SL(2,\R)}={3k \over k-2}$.  Expanding the currents in modes as
\begin{equation}
\label{modos}
J^{3,\pm}(z)= \sum_{n \in \mathbb{Z}} {J^{3,\pm}_n \over z^{n+1}}\,,
\end{equation}
the current algebra~(\ref{opeads})\ can be written as
\begin{eqnarray}
[J^3_n, J^3_m ]    &=& - {k \over 2} n \delta_{n+m,0} \,,
\nonumber\\{}
[J^3_n, J^\pm_m ]  &=& \pm J^\pm_{n+m} \,,
\nonumber\\{}
[J^+_n , J^-_m ]   &=& -2J^3_{n+m} + kn\delta_{n+m,0}\,,
\label{comm}
\end{eqnarray}
with all other commutators vanishing.

The highest weight representations of~(\ref{opeads})\
and~(\ref{comm}) are built starting from unitary representations of
the $\SL(2,\R)$ Lie algebra $J^{3,\pm}_0$.  These are characterized by
the eigenvalues of the Casimir operator $\frac12 (J^+_0 J^-_0 + J^-_0
J^+_0 ) - (J^3_0)^2$, which will be denoted by $-j(j-1)$, and the
states within each representation are labelled by the eigenvalue of
$J^3_0$, which will be denoted by $m$.  These representations are the
primary states for the current algebra.  They are annihilated by
$J^{3,\pm}_{n>0}$, and the Fock space of states is built by acting on
them with $J^{3,\pm}_{n<0}$.

The unitary representations of the $\SL(2,\R)$ Lie algebra
$J^{3,\pm}_0$ which appear in the spectrum of strings moving in an
$AdS_3 \times \cN$ background are~\cite{Malda, Malda2}:
\begin{enumerate}
\item \emph{Highest weight discrete representations}
\[
\cD{j}^+ = \left\{ |j;m \rangle : m=j,j+1, j+2, \ldots \right\}
\]
where $|j;j \rangle$ is annihilated by $J_0^-$ and $j$ is a real
number such that $1/2 < j < (k-1)/2$.

\item \emph{Lowest weight discrete representations}
\[
\cD{j}^- = \left\{ |j;m \rangle : m=-j,-j-1, -j-2, \ldots \right\}
\]
where $|j;-j \rangle$ is annihilated by $J_0^+$ and $j$ is a real
number such that $1/2 < j < (k-1)/2$.

\item \emph{Continuous representations}
\[
\cCo{j,\alpha} = \left\{ |j,\alpha;m \rangle : m= \alpha, \alpha \pm
1,\alpha \pm 2, \ldots \right\}
\]
where $0 \leq \alpha < 1$ and $j = 1/2 + i s $, where $s$ is a real
number.
\end{enumerate}

The bounds on $j$ appearing for $\cD{j}^{\pm}$ can be understood in
terms of consistency conditions for the primary states.  The lower
bound, $ 1/2 < j$, is necessary for the normalizability of the primary
states when their norm is interpreted as the ${\cal L}^2$ inner
product of functions in the $\SL(2,\R)$ group manifold~\cite{DVV}.
Regarding the upper bound, $ j < (k-1)/2$, in~\cite{Giveon} it was
noted that it is necessary for the unitarity of the primary states,
when their norm is interpreted as the two-point function of the vertex
operators creating them from the vacuum.  Moreover, adopting either
the upper or the lower bound for $j$, the other one appears when
imposing the $w=\pm 1$ spectral flow (see below) to be a symmetry of
the spectrum.  Finally, the compelling evidence for the correctness of
these bounds on $j$ comes from the fact that only this range of $j$
appears in the spectrum of the thermal partition function of the
model, computed by path integral techniques in~\cite{Malda2}.

In the computations of section~\ref{section4}, we will be interested
in the characters of the representations~\cite{Malda}, which are very
easy to obtain because the action of the $J_{n<0}^{3,\pm}$ modes is
free for the representations considered~\cite{Kac:fz}.  For $\cD{j}^+$
we have
\begin{eqnarray}
\chi_j^+(q,z)=\tr_{\cD{j}^+} \left( q^{L_0} z^{J_0^3} \right) &= & {
q^{- {j(j-1) \over k-2}} z^j \over
\prod^{\infty}_{n=1}(1-q^n)(1-q^{n-1}z)(1-q^nz^{-1})} \,,\qquad
\nonumber\\
&= & { q^{- {j(j-1) \over k-2}+ \frac18} z^{j - \frac12} \over
i\Theta_1(q,z)}\,.
\label{card}
\end{eqnarray}
Eq.~(\ref{card})\ should be understood as a formal series expansion
which converges to~(\ref{card})\ when $|q|<|z|<1$. The elliptic
function $\Theta_1(q,z)$ is~\cite{mumford}
\begin{eqnarray}
\Theta_1(q,z)&= & -i z^{-1/2}q^{1/8}
\prod_{n=1}^{\infty}(1-q^n)(1-q^{n-1}z)(1-q^nz^{-1})\,,
\nonumber\\
&= & -i \sum_{n \in \mathbb{Z}} (-1)^n z^{n-1/2} q^{{(n-1/2)^2 \over
2}}\,.
\label{thetauno}
\end{eqnarray}
For the $\cC_{j,\alpha}$ representations, the character is
\begin{equation}
\label{carc}
\hat{\chi}_{j=1/2+is,\alpha}(q,z)=\tr_{\cC_{j=1/2+is,\alpha}} \left(
q^{L_0} z^{J_0^3} \right)= \, {q^{s^2 + 1/4 \over k-2} \over
\prod^{\infty}_{n=1}(1-q^n)^3} \sum_{n \in \mathbb{Z}}z^{\alpha+n} \,.
\end{equation}
The character of $\cD{j}^-$ can be obtained from the spectral flow of
$\cD{j}^+$ (see below).

\subsection{$\SL(2,\R)/\U(1) \times \U(1)$ decomposition and spectral flow}\label{section2.2}

It will be convenient to decompose both the currents and the spectrum
into the direct product of a parafermionic $\SL(2,\R)/\U(1)$ model and
the timelike $\U(1)$ model corresponding to $J^3(z)$.  For the
currents $J^{3,\pm}(z)$, this can be performed by representing them~as
\begin{eqnarray}
J^3(z) &= & - \sqrt{\frac{k}2} \d X(z) \,,
\nonumber\\
J^{\pm}(z) &= & \psi^{\pm}(z) e^{\pm \sqrt{\frac2k} X} \,,
\label{decomp}
\end{eqnarray}
with
\begin{eqnarray}
X(z)X(w) &\sim &  -\log (z-w) \,,
\nonumber\\
X(z) \psi^{\pm}(w) &\sim & \, 0 \,.
\label{equis}
\end{eqnarray}
Note that the field $X(z)$ is antihermitean, in order for $J^3(z)$ to
be hermitean and $J^+(z)^{\dagger}=J^-(z)$.  The fields $\psi^{\pm}$
form an $\SL(2,\R)/\U(1)$ parafermionic theory introduced
in~\cite{Lykken}, and are a non-minimal generalization of the
$\SU(2)/\U(1)$ parafermions of~\cite{Fateev}.  The OPEs of
$\psi^{\pm}$ can be read from~(\ref{opeads})\ and~(\ref{equis}).

The stress tensor is similarly decomposed into
\begin{equation}
\label{stressdeco}
T^{{\SL(2,\R)}} = T^{{\SL(2,\R)/\U(1)}} + T^{{\U(1)}}\,,
\end{equation}
with
\begin{equation}
\label{stressuno}
T^{{\U(1)}}= - \frac1k J^3J^3 \,.
\end{equation}
The decomposition can also be performed on the representations.
Indeed, every representation $\cD{j}^+, \cCo{j}$ gives rise to an
infinite number of parafermionic representations
$\lambda_{j,n},\hat{\lambda}_{j,n}$ with $n \in \hZ$.  The $\cD{j}^+,
\cCo{j}$ Hilbert spaces can be decomposed into
\begin{eqnarray}
\cD{j}^{+} & =& \sum_{n \in \mathbb{Z}} \lambda_{j,n} \otimes \cU{j+n}\,,
\nonumber\\
\cCo{j,\alpha} & =& \sum_{n \in \mathbb{Z}} \hat{\lambda}_{j,\alpha+n}
\otimes \cU{\alpha+n} \,,
\label{decompo}
\end{eqnarray}
where $\cU{m}$ is a highest-weight representation of the timelike
current $J^3(z)$ with $J^3_0$ eigenvalue~$m$. The
$\lambda_{j,n},\hat{\lambda}_{j,n}$ representations are built by
acting with the modes of $\psi^{\pm}$ on para\-fer\-mio\-nic
highest-weight states.  The reader can find details
in~\cite{Lykken,Fateev,cha,sfetsos}.  We will be mainly interested in
their characters~\cite{cha,sfetsos}, which are
\begin{eqnarray}
\lambda_{j,n}(q) & =&\tr_{\lambda_{j,n}} \!\! \lp q^{L_0} \rp = {q^{-
{j(j-1) \over k-2} + \frac{(j+n)^2}k} \over
\prod_{n=1}^{\infty}(1-q^n)^2} \sum^{\infty}_{s=0}(-1)^s q^{\frac12
s(s+2n+1) } \,,
\nonumber\\
\hat{\lambda}_{j,\alpha+n}(q)&=& \tr_{\hat{\lambda}_{j,\alpha+n}} \!\!
\lp q^{L_0} \rp = {q^{- {j(j-1) \over k-2} + \frac{(\alpha+n)^2}k}
\over \prod_{n=1}^{\infty}(1-q^n)^2} \,.
\label{carpf}
\end{eqnarray}
In appendix~\ref{section6} we show explicitly that the
decompositions~(\ref{decompo})\ are, in terms of the characters,
\begin{eqnarray}
\chi^+_j(q,z) & =& \sum_{n  \in \mathbb{Z}} z^{j+n} \,
\lambda_{j,n}(q)\, \zeta_{j+n}(q) \,,
\nonumber\\
\hat{\chi}_{j=1/2+is,\alpha}(q,z) & =& \sum_{n \in \mathbb{Z}}
z^{\alpha +n} \, \hat{\lambda}_{j,\alpha+n}(q)\, \zeta_{\alpha+n}(q)\,,
\label{cardeco}
\end{eqnarray}
where
\begin{equation}
\label{pfuno}
\zeta_{m}(q)\equiv \tr_{\cU{m}} \left( q^{L_0} \right) = {q^{-{m^2
\over k}} \over \prod^{\infty}_{n=1}(1-q^n)} \,.
\end{equation}
Now, given any integer $w$, the algebra~(\ref{comm})\ is preserved by
the spectral flow $J_n^{3,\pm} \rightarrow \tilde{J}_n^{3,\pm}$ first
introduced in~\cite{Henningson:1991jc} and defined by
\begin{eqnarray}
\tilde{J}_n^3 & =& J_n^3 - {k \over 2} w \delta_{n,0}  \,,
\nonumber\\
\tilde{J}_n^\pm & =& J_{n \pm w}^\pm \,.
\label{flow}
\end{eqnarray}
This transformation
maps  $L_0$ to
\begin{equation}
\label{flowele}
\tilde{L}_0= L_0 + wJ^3_0 - \frac{k}4w^2 \,.
\end{equation}
The spectral flow can be implemented in~(\ref{decomp})\ by
\begin{equation}
\label{flowx} X(z) \longrightarrow \tilde{X}(z)=X(z) +
w \sqrt{\frac{k}{2}} \log z \,,
\end{equation}
and leaving the parafermions $\psi^{\pm}(z)$ untouched. For $w=\pm 1$,
the symmetry~(\ref{flow})\ maps the $\cD{j}^{\mp}$ representation into
$\cD{k/2-j}^{\pm}$. Note that under this mapping the upper and lower
bounds of $j$ are interchanged. Including the $w=\pm1$ spectrally
flowed states, we can just consider $\cD{j}^{+}$ representations.

For generic $w$, it was shown in~\cite{Malda,Malda2} that the Hilbert
space $\cD{j}^+$ is mapped into new representations which must be
included in a consistent description of a string propagating in $AdS_3
\times \cN$. The way to include these representations is to take usual
representations $\cD{j}^+, \cCo{j,\alpha}$ for flowed operators
$\tilde{J}^{3,\pm}_n$, and measure their quantum numbers with unflowed
$J^3_0,L_0$.  Alternatively, instead of this \emph{active} procedure,
we can use a \emph{passive} one, and obtain the flowed sectors by
acting with the unflowed operators $J^{3,\pm}_n$ on flowed
primaries. Starting with an $\SL(2,\R)$ vertex operator
\begin{equation}
\label{vertex}
\Phi_{j,m}= \Psi_{j,m}  e^{m \sqrt{\frac2{k}}X} \,,
\end{equation}
where $\Psi_{j,m}$ is a parafermionic primary, the flowed vertex
operator is obtained applying to it the mapping~(\ref{flowx}),
yielding
\begin{equation}
\label{vertexflow}
\Phi_{j,m}^{w}= \Psi_{j,m} e^{(m+wk/2) \sqrt{\frac2{k}}X} \,.
\end{equation}
Note that for $\cCo{j, \alpha}$ representations, all the excited
physical states belong to flowed sectors.  This can be seen from the
mass-shell condition,
\begin{equation}
\label{capademasa}
L_0={\frac14 + s^2 \over k-2} -w(\alpha + n) + \frac{kw^2}{4} + N +
h^{\cN} - 1 =0 \,,
\end{equation}
where $h^{\cN}$ is a highest weight of the inner theory and N is the
level. It is clear that for $w=0$, eq.~(\ref{capademasa})\ can only be
satisfied if $N=0$.

The characters of the flowed $\cD{j}^w$ representations are
\begin{eqnarray}
\chi^w_j(q,z)&= & \tr_{\cD{j}^w} \left( q^{L_0} z^{J_0^3} \right) =
\tr_{\cD{j}^+} \left( q^{\tilde{L}_0-w\tilde{J}^3_0-kw^2/4}
z^{\tilde{J}_0^3+kw/2} \right)
\nonumber\\
&=& q^{-kw^2/4} z^{kw/2}  \tr_{\cD{j}^+} \left( q^{\tilde{L}_0}
(zq^{-w})^{\tilde{J}_0^3} \right)
\nonumber\\
&= & { (-1)^w q^{\left[- {j(j-1) \over k-2} - (k-2)\frac{\,w^2}{4}
-(2j -1)\frac{w}{2} + \frac18 \right]}  z^{\left[j +
(k-2)\frac{w}{2} - \frac12\right]} \over i\Theta_1(q,z)} \,.
\label{cardf}
\end{eqnarray}
where we have used the property~\cite{mumford}
\begin{equation}
\label{pete}
\Theta_1(q,zq^{-w})=(-1)^w q^{-w^2/2}z^w \Theta_1(q,z) \,.
\end{equation}
For the $\cCo{j,\alpha}^w$ representations we obtain similarly
\begin{eqnarray}
\hat{\chi}^w_{j=1/2+is,\alpha}(q,z)&= &\tr_{\cCo{j,\alpha}^w} \left(
q^{L_0} z^{J_0^3} \right)
\nonumber\\
&= & {q^{{s^2 + 1/4 \over k-2}-kw^2/4 -w\alpha} \over
\prod^{\infty}_{n=1}(1-q^n)^3} \sum_{n \in \mathbb{Z}}z^{\alpha+n+
kw/2}q^{-wn} \,.\qquad
\label{carcf}
\end{eqnarray}
As is clear from~(\ref{vertexflow}), the effect of the spectral flow
is to ``dislocate" the parafermionic and the $J^3_0$ quantum numbers by
an amount $wk/2$.  The decompositions~(\ref{decompo})\ are generalized~to
\begin{eqnarray}
\cD{j}^{w} & =& \sum_{n \in \mathbb{Z}} \lambda_{j,n} \otimes
\cU{j+n+kw/2} \,,
\nonumber\\
\cCo{j,\alpha}^{w} & =& \sum_{n \in \mathbb{Z}}
\hat{\lambda}_{j,\alpha+n} \otimes \cU{\alpha+n+kw/2} \,.
\label{decompoflow}
\end{eqnarray}
In appendix~\ref{section8} we show explicitly that, in terms of the
characters, the decompositions~(\ref{decompoflow})\ are
\begin{eqnarray}
\chi^w_j(q,z) & =& \sum_{n \in \mathbb{Z}} z^{j+n+ kw/2} \,
\lambda_{j,n}(q)\, \zeta_{j+n+kw/2}(q) \,,
\nonumber\\
\hat{\chi}^w_{j=1/2+is,\alpha}(q,z) & =& \sum_{n \in \mathbb{Z}}
z^{\alpha+n+ kw/2} \hat{\lambda}_{j,\alpha+n}(q)\,
\zeta_{\alpha+n+kw/2}(q) \,.
\label{cardecof}
\end{eqnarray}

\subsection{The supersymmetric algebra}\label{section2.3}

For fermionic strings in $AdS_3 \times \cN$, we should consider the
current algebra of affine supersymmetric $\SL(2,\R)$ at level k. It is
generated by three dimension-$\frac12$ supercurrents $\psi^{3,\pm}+
\theta J^{3,\pm}$, whose modes satisfy
\begin{eqnarray}
[J^3_n, J^3_m ] &=& - {k \over 2} n \delta_{n+m,0} \,,
\nonumber\\{}
[J^3_n, J^\pm_m ] &=& \pm J^\pm_{n+m} \,,
\nonumber\\{}
[J^+_n , J^-_m ] &=& -2J^3_{n+m} + kn\delta_{n+m,0} \,,
\nonumber\\{}
[J^3_n, \psi^\pm_m]&=&\pm \psi^\pm_{n+m} \,,
\nonumber\\{}
[J^\pm_n,\psi^\mp_m]&=&\mp 2\psi^3_{n+m} \,,
\nonumber\\{}
[J^\pm_n, \psi^3_m]&=&\mp \psi^\pm_{n+m} \,,
\nonumber\\
\{ \psi^3_n,\psi^3_m \}&=& -{k \over 2} \delta_{n+m,0}\,,
\nonumber\\
\{ \psi^+_n,\psi^-_m \}&=& k \delta_{n+m,0} \,,
\label{commfer}
\end{eqnarray}
with all other (anti)commutators vanishing.  The modding of $J^a_n$ is
integer and that of $\psi^a_n$ half-integer (Neveu-Schwarz) or integer
(Ramond).  The worldsheet CFT is supersymmetric, the Sugawara stress
tensor and the supercurrent being
\begin{eqnarray}
T^{\SL(2,\R)} &=&{1 \over 2k}[j^+j^- +j^-j^+] - {1 \over k}j^3j^3 -{1
\over 2k}[\psi^+ \partial \psi^- + \psi^- \partial \psi^+] +{1 \over
k}\psi^3 \partial \psi^3 \,,
\nonumber\\
G^{\SL(2,\R)} &=&{1 \over k} [\psi^+j^- + \psi^-j^+] -{2 \over
k}\psi^3j^3 - {2 \over k^2} \psi^+\psi^-\psi^3 \,.
\label{tensorT}
\end{eqnarray}
As usual, purely bosonic currents can be defined by
\begin{eqnarray}
j^a &=& J^a - \hat{J}^a \,,
\nonumber\\
\hat{J}^a &=& -{i \over k} f^a_{~bc} \psi^b \psi^c \,.
\label{bosonicJ}
\end{eqnarray}
The $j^a$ form an affine $\SL(2,\R)$ bosonic algebra at level $k+2$.
The $\hat{J}^a$ and the $\psi^a$ form a supersymmetric affine
$\SL(2,\R)$ algebra at level $-2$ which commutes with $j^a$. The
spectrum of the theory is the direct product of the Hilbert spaces of
both theories.  In the bosonic sector, the reprentations are the ones
considered above.  For the $\psi^a$ currents, we have the usual
representations for free fermions, for both Neveu-Schwarz and Ramond
sectors. The spectral flow can be also defined in the
$\psi^a,\hat{J}^a$ theory, but it just amounts to a rearrangement of
the usual spectrum, as shown in~\cite{Pakman:2003cu}.

\section{The Hodge theory of the BRST complex}\label{section3}

The formalism reviewed below was developed in~\cite{josebrst}, and can
be found in~\cite{jose}.  For similar results in the context of
compact Lie algebras, see~\cite{vanHolten:1990xd}.

We will denote by $\cF$ the Hilbert space composed of the ghost plus
matter sectors, with the restriction $b_0=L_0=0$, and also
$\beta_0=G_0=0$ for fermionic strings in the Ramond sector.  Although
$\cF$ is infinite dimensional (due to the infinite degeneracy of the
base states of the $\SL(2,\R)$ representations), it can be decomposed
into
\begin{equation}
\label{efesuma}
\cF = \bigoplus_{m} \cF(m) \,,
\end{equation}
where each subspace $\cF (m)$ has $J^3_0=m$ and is finite dimensional.
In the following, we will consider each $\cF(m)$ space separately,
although we will keep the notation $\cF$.

The space $\cF$ has a natural grading under $N_g$ as
\begin{equation}
\label{sumag}
\cF = \bigoplus_{g \in \mathbb{Z}} \cF_g \,,
\end{equation}
where $\cF_g$ has ghost number $g$.  The action induced by $Q$ in each
$\cF_g$ will de denoted by
\begin{equation}
\label{qun}
Q_g: \cF_g \rightarrow \cF_{g+1} \,,
\end{equation}
and the following sequence is a graded complex
\begin{equation}
\label{secuencia}
\cdots {\longrightarrow} \cF_{g-1} \stackrel{Q_{g-1}}{\longrightarrow} \cF_{g}
\stackrel{Q_{g}}{\longrightarrow}\cF_{g+1}{\longrightarrow} \cdots\,.
\end{equation}
For each $g$ we define the kernel and the image of $Q$ as
\begin{eqnarray}
\rker Q_g & =& \{\xi \in \cF_g \, | \, Q \xi =0 \} \,,
\nonumber\\
\rim Q_g & =& \{ Q\xi \, | \, \xi \in \cF_g \} \,.
\label{kerima}
\end{eqnarray}
The cohomology of $Q$ in $\cF$ is then defined as
\begin{equation}
\label{defcoho}
\rH= \bigoplus_g \rH_g \,,
\end{equation}
with
\begin{equation}
\label{defcohog}
\rH_g = {\rker Q_g \over \rim Q_{g-1}} \,.
\end{equation}
Since the indefinite inner product in $\cF$ is non-degenerate, an
orthogonal basis in $\cF$ can be constructed.  Let us define the
action of a linear hermitean operator $\cC$ on the states of such a
basis as $1$ $(-1)$ if the state has positive (negative) norm.  Then
for any pair of states $\xi, \eta \in \cF$ the inner product
\begin{equation}
\label{inprodc}
\inc{\xi}{\eta} \equiv \inn{\xi}{\cC \eta} \,,
\end{equation}
is clearly positive definite.  The latter implies that $\cC$ maps
$\cF_g$ into $\cF_{-g}$ because the antihermitean character of $N_g$
implies that two states can have nonzero $\inn{}{}$ inner product only
if they have opposite ghost number.  Indeed, $\cC$ interchanges the
$b_n$ and the $c_n$ modes of any state (and the $\beta_n$ and
$\gamma_n$ for fermionic strings), as we show explicitly in
appendix~\ref{section7}. Moreover, this mapping from $\cF_g$ to
$\cF_{-g}$ is a linear isomorphism because each $\cF_g$ is finite
dimensional and $\cC$ has empty kernel because $\cC^2=1$. Note that
the action of $\cC$ is similar to the action of the Hodge $*$ operator
on p-forms.

Under this new inner product $Q$ is no longer hermitean, but its
conjugate $Q^*$ can be obtained from
\begin{equation}
\label{qconj}
\inc{\xi}{Q\eta} = \inn{\xi}{\cC Q \eta} = \inn{\cC Q\cC \xi}{\cC
\eta} = \inc{Q^*\xi}{\eta} \,,
\end{equation}
where we have used $\cC^2=1$, so we have $Q^*= \cC Q \cC$. From its
definition it is clear that $\left(Q^* \right)^2 = 0 $ and that $Q^*$
carries ghost number $-1$ and is hermitean under $\inn{}{}$.  As for
$Q$, we will denote by $Q^*_g$ the action of $Q^*$ on $\cF_g$.

The positive definiteness of~(\ref{inprodc})\ allows us to decompose
$\cF_g$ both as
\begin{equation}
\label{decouno}
\cF_g = \rim Q_{g-1} \oplus ( \rim Q_{g-1} )^{\perp} \,,
\end{equation}
and as
\begin{equation}
\label{decodos}
\cF_g = \rim Q^*_{g+1} \oplus ( \rim Q^*_{g+1} )^{\perp} \,,
\end{equation}
where $\lp \cdots \rp ^{\perp}$ means orthogonal with respect to $\inc{}{}$.
But, as can be easily checked, we have $( \rim Q^*_{g+1} )^{\perp} = \rker Q_{g}$
and $( \rim Q_{g-1} )^{\perp} = \rker Q^*_{g}$.
Applying now the decomposition~(\ref{decouno})\ to  $\rker Q_{g}$ we have
\begin{equation}
\label{decoker}
\rker Q_{g} = \rim Q_{g-1} \oplus \cH_g \,,
\end{equation}
where we have defined
\begin{equation}
\label{hache}
\cH_g = \rker Q_g \cap \rker Q^*_g \,.
\end{equation}
The elements of $\cH_g$ provide us with a unique representative for
each cohomology class $\textrm{H}_g$. Indeed, from~(\ref{hache})\ we
see that the elements of $\cH_g$ are annihilated by $Q_g$, and two
representatives of the same class of $\textrm{H}_g$ always differ by
an element belonging entirely to the first term
of~(\ref{decoker}). Moreover, it is easy to check (using that
$\inc{}{}$ is positive definite) that a state belongs to $\cH_g$ iff
it is \emph{harmonic}, i.e., it is annihilated by the \emph{laplacian}
\begin{equation}
\label{laplace}
\Delta \equiv Q Q^* + Q^* Q \,.
\end{equation}
Inserting now~(\ref{decoker})\ into~(\ref{decodos})\ we obtain the
\emph{Hodge decomposition}
\begin{equation}
\label{hodge}
\cF_g = \rim Q_{g-1} \oplus \rim Q^*_{g+1} \oplus \cH_g \,.
\end{equation}
It is easy to check that the isomorphism between $\cF_g$ and
$\cF_{-g}$ provided by $\cC$ acts on the terms of~(\ref{hodge})\ as
\begin{eqnarray}
\cC : & & \rim Q_g \longleftrightarrow \rim Q^*_{-g} \,,
\nonumber\\
\cC : & & \cH_g \longleftrightarrow \cH_{-g} \,,
\label{mapeos}
\end{eqnarray}
and from the second line of~(\ref{mapeos})\ we find that the
cohomologies $\textrm{H}_g$ and $\textrm{H}_{-g}$ are isomorphic, a
result known as \emph{Poincar\'e duality}.

Now, it is clear that
\begin{equation}
\label{dim}
\di \cF_g = \di \rker Q_g + \di \rim Q_g \,,
\end{equation}
and from~(\ref{decoker})\ we have $\di \rker Q_g = \di \rim Q_{g-1} +
\di \cH_g$, which implies
\begin{equation}
\label{dimdos}
\di \cF_g = \di \cH_g + \di \rim Q_g + \di \rim Q_{g-1} \,.
\end{equation}
Performing an alternating sum of~(\ref{dimdos})\ over all $g$, the
last two terms cancel pairwise and we obtain the
\emph{Euler-Poincar\'e identity}:
\begin{equation}
\label{eulerpoincare}
\sum_g (-1)^g \di \cF_g = \sum_g (-1)^g \di \cH_g \,\,.
\end{equation}

\subsection{The FGZ proof of the no-ghost theorem}\label{section3.1}

The physical space of states is defined to be the $Q$ cohomology in
$\cF$.  Since the latter is concentrated at zero ghost number, it can
be identified with $\cH_0$, and the no-ghost theorem asserts that
states in $\cH_0$ have positive norm under the usual inner product
$\inn{}{}$.

Note that from the second line of~(\ref{mapeos})\ we see that $\cC$
maps $\cH_0$ to itself, so it can be diagonalized in $\cH_0$ with
eigenvalues $+1$ or $-1$ (because $\cC^2=1$), corresponding to states
with positive or negative $\inn{}{}$ norm, respectively.  Calling
$N_+$ $(N_-)$ to the number of $+1$ $(-1)$ eigenvalues, we have
\begin{equation}
\label{des}
\di \cH_0 \geq N_+ - N_- = \tr_{\cH_0}\cC \,,
\end{equation}
and it is clear that the no-ghost theorem is proved iff the equality
in~(\ref{des})\ holds.

The centrality of the vanishing theorem comes now to the fore, since
having $\textrm{H}_{g \neq 0}= 0$ implies, through the
Euler-Poincar\'e identity~(\ref{eulerpoincare}),
\begin{equation}
\label{epdual}
\ind\equiv \sum_{g} (-1)^g \di \cF_g = \sum_{g} (-1)^g \di \cH_g = \di
\cH_0 \,,
\end{equation}
where $\ind$ is the \emph{Euler index} of the BRST complex.  As for
$\tr_{\cH_0}\cC$, notice that the isomorphism~(\ref{mapeos})\ under
$\cC$ between $\rim Q_{-1}$ and $\rim Q^*_1$ implies
\begin{equation}
\label{trazanula}
\tr_{ { \{ \rim Q_{-1} \oplus \rim Q^*_1 \} } }  \cC = 0 \,,
\end{equation}
so from~(\ref{hodge})\ we have
\begin{equation}
\label{trazaefecero}
\tr_{\cH_0}\cC = \tr_{\cF_0}\cC \,.
\end{equation}
Similarly, the isomorphism under $\cC$ between $\cF_g$ and $\cF_{-g}$
implies for $g \neq 0$ that
\begin{equation}
\label{trazanulados}
\tr_{{ \{ \cF_{g} \oplus \cF_{-g} \} }} \, \cC = 0 \,,
\end{equation}
and from this we obtain
\begin{equation}
\label{sign}
\sig \equiv \tr_{\cF}\, \cC = \tr_{{{\bigoplus}_g \cF_g}} \, \cC =
\tr_{\cH_0}\cC \,,
\end{equation}
where $\sig$ is the \emph{signature} of the BRST
complex. Putting~(\ref{des})--(\ref{sign})\ together, the no-ghost
theorem is equivalent to the identity
\begin{equation}
\label{nog}
\sig = \ind\,.
\end{equation}

\section{No-ghost theorem for strings in $AdS_3$}\label{section4}

We will prove the identity~(\ref{nog})\ by comparing
\begin{eqnarray}
\sig(q,z) &=  &  \tr \left(\cC q^{L_0} z^{J_0^3}\right) ,
\nonumber\\
\ind(q,t)  &=  & \tr \left((-1)^{N_g} q^{L_0} z^{J_0^3}\right) ,
\label{see}
\end{eqnarray}
where the traces will be taken over the different representations.
Note that for our purposes the formal expansions in $z$ and $q$ are of
very different nature.  Because of the mass shell condition $L_0=0$,
in the expansion in $q$ we are only interested in the term independent
of $q$.\footnote{Note that the condition $b_0=0$ (and $\beta_0=0$) is
already taken into account by just ignoring the degeneracy of the
ghost vacuum in the traces~(\ref{see}).}  On the other hand, we need
all the powers of $z$, since the decomposition~(\ref{efesuma})\ was
only done in order to separate $\cF$ into finite-dimensional
subspaces.

In the computation of~(\ref{see}), we will use the fact that the trace
is multiplicative under tensor products and additive under direct
sums.  We will consider representations with arbitrary flow parameter
$w$, which of course include the regular representations ($w=0$).

\subsection{Bosonic strings}\label{section4.1}

In~\cite{DPL} it was shown that $\SL(2,\R)/\U(1)$ representations
coming from $\cCo{j,\alpha}$ representations with $k>2$, and those
coming from $\cD{j}$ with $k>2, j < k/2$, are unitary.  Since these
bounds are satisfied in our case, the only source of negative norms
are the modes of $J^3(z)$.  The action of $\cC$ in the $\SL(2,\R)$
sector should clearly be
\begin{eqnarray}
\cC \psi^{\pm}(z) \cC & =& \psi^{\pm}(z) \,,
\nonumber\\
\cC J^3_{-n} \cC & =& -J^3_{-n} \,.
\label{accionc}
\end{eqnarray}
We decompose $\cF$ into
\begin{equation}
\label{decfb}
\cF=\cF_{(b,c)} \otimes \cF_{\SL(2,\R)} \otimes \cF_{\cN} \,,
\end{equation}
and we further decompose the representations of $\cF_{\SL(2,\R)}$ as
in~(\ref{decompoflow})--(\ref{cardecof}).

The signature of the $\cD{j}^w$ representations can now be computed as
\begin{equation}
\label{signatura}
\sig_{\cD{j}^w} (q,z) =  \sig_{(b,c)}(q) \times \sum_{n \in
\mathbb{Z}} z^{j+n+kw/2}  \sig_{\lambda_{j,n}}(q)
\sig_{\cU{j+n+kw/2}}(q) \times \sig_{\cN}(q) \,.
\end{equation}
The traces in~(\ref{signatura})\ can be computed as follows.  In the
ghost sector, using the decomposition~(\ref{ghostdecom}), we have
\begin{equation}
\label{signabc}
\sig_{(b,c)}(q) = \tr_{(b,c)} \left(\cC q^{L_0} \right)= q^{-1}
\prod_{n=1}^{\infty}(1-q^n)(1+q^n) \,.
\end{equation}
The signature of the timelike $\U(1)$ is easily found
from~(\ref{accionc})\ to be
\begin{equation}
\label{signau}
\sig_{\cU{j+n+kw/2}}(q)=\tr_{{\cal U}_{j+n+kw/2}} \lp {\cal C} q^{L_0}
\rp = {q^{-{(j+n+kw/2)^2 \over k}} \over \prod_{n=1}^{\infty}(1+q^n)}
\,.
\end{equation}
Since the theories $\SL(2,\R)/\U(1)$ and $\cN$ are unitary, their
signature is their character.  So
from~(\ref{signatura})--(\ref{signau})\ the total signature is then
\begin{equation}
\label{signais}
\sig_{\cD{j}^w} (q,z) = q^{-1} \prod^{\infty}_{n=1}(1-q^n) \times
\sum_{n \in \mathbb{Z}} q^{{(j+n+kw/2) \over k}^2} z^{j+n+kw/2}
\lambda_{j,n}(q) \times \trk \,.
\end{equation}
The index of the $\cD{j}^w$ representation is
\begin{equation}
\label{index}
\ind_{{\cD{j}^w}} (q,t) = \ind_{(b,c)}(q) \tr_{\cD{j}^w} \left(
q^{L_0} z^{J_0^3} \right)  \trk \,.
\end{equation}
In the ghost sector, decomposing $\cF_{(b,c)}$ as
\begin{equation}
\label{ghostdecomdos}
\cF_{(b,c)} = \bigotimes_{n=1}^{\infty} \left\{ |0 \rangle_g, b_{-n}|0
\rangle_g \right\} \otimes \{ |0 \rangle_g, c_{-n}|0 \rangle_g \} \,,
\end{equation}
we have
\begin{equation}
\label{indexbc}
\ind_{(b,c)}(q)=\tr_{(b,c)} \left((-1)^{N_g} q^{L_0} \right) = q^{-1}
\prod^{\infty}_{n=1}(1-q^n)^2 \,,
\end{equation}
and the index can be written as
\begin{equation}
\label{indexis}
\ind_{{\cD{j}^w}}= q^{-1} \prod^{\infty}_{n=1}(1-q^n)^2 \times
\chi^w_j(q,z)\times \trk \,.
\end{equation}
Finally, the equality between~(\ref{signais})\ and~(\ref{indexis})\
follows immediately from~(\ref{pfuno})\ and~(\ref{cardecof}).

Proceeding similarly for the $\cC_{j,\alpha}^w$ representations, we obtain
\begin{eqnarray}
\sig_{\cC_{j,\alpha}^w}& =& q^{-1} \prod^{\infty}_{n=1}(1-q^n) \times
\sum^{\infty}_{n=-\infty} q^{{(\alpha+n+kw/2) \over k}^2}
z^{\alpha+n+kw/2} \hat{\lambda}_{j,\alpha+n}(q) \times \trk \,,
\nonumber\\
\ind_{\cC_{j,\alpha}^w}& =& q^{-1} \prod^{\infty}_{n=1}(1-q^n)^2
\times \hat{\chi}^w_{j,\alpha}(q,z)\times \trk \,,
\label{sigcon}
\end{eqnarray}
and the equality follows again from~(\ref{pfuno})\ and~(\ref{cardecof}).

\subsection{Fermionic strings}\label{section4.2}

For NSR strings, the computation of the traces is greatly simplified
by decomposing the $AdS_3$ sector at level $k$ into a purely bosonic
$\SL(2,\R)$ model at level $k+2$ and three free fermions $\psi^a$
($a=1,2,3$) (See section~\ref{section2.3}).  From~(\ref{commfer}), we
see that the action of $\cC$ on the free fermions~is
\begin{equation}
\label{accioncfer}
\cC \psi^{a}_n \cC = (-1)^{\delta_{a,3}} \, \psi^{a}_n(z) \,.
\end{equation}
The total Hilbert space $\cF$ can then be expressed as
\begin{equation}
\label{decff}
\cF=\cF_{(b,c)} \otimes \cF_{(\beta,\gamma)} \otimes \cF_{\SL(2,\R)}
\otimes \cF_{\psi^a} \otimes \cF_{\cN} \,.
\end{equation}
Note that the traces on $\cF_{(b,c)} \otimes \cF_{\SL(2,\R)} \otimes
\cF_{\cN}$ can be taken from the above bosonic computation, including
there the fermionic sectors of $\cN$.  We will compute the traces for
$\cD{j}^w$ representations, the $\cCo{j,\alpha}^w$ case being
identical.

\paragraph{Neveu-Schwarz sector.}

Using the decomposition~(\ref{ghdemm}), and~(\ref{accioncfer}), we find
\begin{eqnarray}
\sig_{(\beta,\gamma)}(q)&=&\tr_{(\beta,\gamma)} \left(\cC q^{L_0}
\right)= \prod_{r=\frac12}^{\infty} {q^{\frac12} \over (1-q^r)(1+q^r)}\,,
\label{sigbg}\\
\sig_{\psi^a}(q) &=&\tr_{\psi^a} \left(\cC q^{L_0}
\right)=\prod_{r=\frac12}^{\infty} (1-q^r)(1+q^r)^2 \,.
\label{sigpsi}
\end{eqnarray}
The total signature is then,
\begin{equation}
\label{signs}
\sig^{NS}(q,z)=\sig_{\cD{j}^w}(q,z)  \sig_{(\beta,\gamma)}(q)
\sig_{\psi^a}(q) = q^{\frac12} \times \sig_{\cD{j}^w}(q,z) \times
\prod_{r=\frac12}^{\infty} (1+q^r) \,.
\end{equation}
To find $\ind_{(\beta,\gamma)}$, we decompose
\begin{equation}
\label{bgdecdos}
\cF_{(\beta,\gamma)} = \bigotimes_{r=1/2}^{\infty}
\bigoplus_{N_r=0}^{\infty} \left\{ \beta^{N_r}_{-r}|0 \rangle_g
\right\} \otimes \bigotimes_{r=1/2}^{\infty}
\bigoplus_{N_r=0}^{\infty} \left\{ \gamma^{N_r}_{-r}|0 \rangle_g
\right\} ,
\end{equation}
and then
\begin{equation}
\label{indbg}
\ind_{(\beta,\gamma)}(q)=\tr_{(\beta,\gamma)} \left((-1)^{N_g} q^{L_0}
\right)=\prod_{r=\frac12}^{\infty} {q^{\frac12} \over(1+q^r)^2} \,.
\end{equation}
We also need
\begin{equation}
\label{indpsi}
\tr_{\psi^a} \left( q^{L_0}\right)= \prod_{r=\frac12}^{\infty}
(1+q^r)^3 \,.
\end{equation}
So the total index is
\begin{equation}
\label{indns}
\ind^{NS}(q,z)=\ind_{\cD{j}^w}(q,z) \ind_{(\beta,\gamma)}(q)
\tr_{\psi^a} \left( q^{L_0}\right) =q^{\frac12} \times
\ind_{\cD{j}^w}(q,z) \times \prod_{r=\frac12}^{\infty}(1+q^r) \,.
\end{equation}
The equality between~(\ref{signs})\ and~(\ref{indns})\ follows then
from the bosonic results.

\paragraph{Ramond sector.}

The traces are computed similarly, giving for the signatures
\begin{eqnarray}
\sig_{(\beta,\gamma)}(q)&=& \prod_{r=1}^{\infty} {q^{3 \over 8} \over
(1-q^r)(1+q^r)} \,,
\label{sigbgr}\\
\sig_{\psi^a}(q)&=&q^{\frac3{16}} \times \prod_{r=1}^{\infty}
(1-q^r)(1+q^r)^2 \,,
\label{sigpsir}
\end{eqnarray}
and
\begin{equation}
\label{sigram}
\sig^{R}(q,z)=\sig_{\cD{j}^w}(q,z) \sig_{(\beta,\gamma)}(q)
\sig_{\psi^a}(q) =q^{9 \over 16} \times \sig_{\cD{j}^w}(q,z) \times
\prod_{r=1}^{\infty} (1+q^r) \,.
\end{equation}
For the computation of the index, we obtain
\begin{eqnarray}
\ind_{(\beta,\gamma)}(q)&=& \prod_{r=1}^{\infty} {q^{3 \over 8}
\over(1+q^r)^2} \,,
\label{indbgr}\\
\tr_{\psi^a} \left( q^{L_0}\right)&=&q^{\frac3{16}} \times
\prod_{r=1}^{\infty} (1+q^r)^3 \,,
\label{indpsir}
\end{eqnarray}
and
\begin{equation}
\label{indram}
\ind^{R}(q,z)=\ind_{\cD{j}^w}(q,z) \, \ind_{(\beta,\gamma)}(q) \,
\tr_{\psi^a} \left( q^{L_0}\right) =q^{9 \over 16} \times
\ind_{\cD{j}^w}(q,z) \times \prod_{r=1}^{\infty}(1+q^r) \,.
\end{equation}
The equality between~(\ref{sigram})\ and~(\ref{indram})\ follows again
from the bosonic results.

Notice that for the computation of $\sig$ in the fermionic case, we
have relied on the unitarity of the bosonic $\SL(2,\R)/\U(1)$ coset,
and we did not need the unitarity of the supersymmetric coset, as in
the OCQ~\cite{Pakman:2003cu}. This is related to the fact that with
the FGZ method we bypass the building of the physical states
themselves, which do belong to the \emph{supersymmetric} coset, modulo
spurious states.

\section{Discussion}\label{section5}

The applicability of the FGZ method depends on two points.  The first
is expressing (a sector of) the matter chiral algebra as a free
$L_-$-Virasoro module. This should be checked in each case, although
the techniques
of~\cite{Asano,Brower,Malda,Hwang:1991an,Evans:1998qu,Pakman:2003cu}
can be adapted or adopted for a wide family of backgrounds.

The second point is having enough control over the representations of
the worldsheet CFT, so as to compute the traces~(\ref{see}).  Based on
the results for $AdS_3 \times \cN$, several related exact CFT
backgrounds could be dealt with successfully. In particular, the BTZ
black hole~\cite{Banados:wn} (an orbifold of $AdS_3$), the non-compact
spaces studied in~\cite{Hwang:1998tr} and pp-waves backgrounds,
obtained as Penrose limits of $AdS_3 \times S^3 \times M^4$
(see~\cite{Hikida:2003fp} and references therein).

An additional class of related backgrounds are gauged WZW models
including an $\SL(2,\R)$ factor, such as the lorentzian 2D black
hole~\cite{Witten:1991yr,DVV} (${\cal M}=\SL(2,\R)/\U(1)$), the
cosmological Nappi-Witten model~\cite{Nappi:1992kv} (${\cal
M}=\SL(2,\R) \times \SU(2) /\U(1)^2$) and the generalization of the
latter studied in~\cite{Elitzur:2002vw}.  These backgrounds are
particularly interesting, because although the matter ${\cal M}$ CFT
(the relative cohomology of the BRST complex in the gauged WZW model)
contains negative-normed states, it cannot be decomposed as $\U(1)
\times {\cal M}/\U(1)$.  Here we can use to advantage the
cohomological analysis of gauged WZW models
in~\cite{jose,Hwang:1993nc}, which show that a vanishing theorem holds
in $G/H$ (for certain subgroups $H$).  This allows to compute $\sig$
and $\ind$ on $G/H$ by taking the traces over $G \otimes \cF_{\rm
ghosts} \otimes H$, and following a chain of identities similar to
those in~(\ref{epdual})\ and~(\ref{sign}).  In this way, as we did in
this work, the flowed sectors of the $\SL(2,\R)$ factor in G can be
included.  This will be worked out in detail in a future work.

\acknowledgments

I thank for discussions G.~d'Appolonio, S.~Elitzur, A.~Giveon, D.~Kazhdan,
B.~Kol,  A.~Konechny, B.~Pioline, P.A.G.~Pisani, A.~Sever and B.~Zwiebach,
and especially \linebreak  J.M.~Figueroa-O'Farrill.
I also thank P.A.G.~Pisani for collaboration at the early stages
of this work.
This work is supported in part by the Israel Academy
of Sciences and Humanities --- Centers of Excellence Program, the
German-Israel Bi-National Science Foundation, and the European RTN
network HPRN-CT-2000-00122.

\paragraph{Note added.}

After this paper appeared in hep-th, the work~\cite{Asano:2003qb}
appeared, where the methods of~\cite{Asano} are applied to $AdS_3
\times \cN$ backgrounds.

\appendix

\section{BRST essentials}\label{section6}

We review here the basics of the $(b,c)$ and $(\beta,\gamma)$ algebra
that we need in this work.  More details can be found
in~\cite{Polchinski:rq}. The zero-mode quantization has some
subtleties, which have been discussed, \emph{inter alia},
in~\cite{henneaux}.

\subsection{Bosonic strings}\label{section6.1}

The anticommuting $(b,c)$ reparametrization ghosts, with conformal
dimensions $(2,-1)$, have the OPEs
\begin{eqnarray}
b(z)c(w)  & \sim& c(z)b(w) \sim  {1 \over z-w} \,,
\nonumber\\
b(z)b(w)  & \sim&  c(z)c(w) \sim \, 0 \,.
\label{opebc}
\end{eqnarray}
The ghost number operator is defined as
\begin{eqnarray}
N_g & =& - \oint dz \, b(z) c(z) \,,
\nonumber\\
& =& \sum_{n=1}^{\infty}\left( c_{-n}b_{n} - b_{-n}c_{n}\right) +
c_0b_0 - \frac12 \,.
\label{ng}
\end{eqnarray}
The ordering constant $-1/2$ can be fixed in several
ways~\cite{Polchinski:rq}, the most immediate one being the
requirement of $N_g$ being antihermitean, i.e., $N_g^\dagger =
-N_g$. For this we use $b_n^{\dagger}=b_{-n}$, $c_n^{\dagger}=c_{-n}$.

The charge of the $(b-c)$ modes under $N_g$ is
\begin{eqnarray}
[N_g,b_m]&=& - b_m \,,
\nonumber\\
{}[N_g,c_m]&=& + c_m \,.
\label{cargasbc}
\end{eqnarray}
The zero modes $b_0$, $c_0$ satisfy the Clifford algebra $\{b_0,c_0
\}=1$, so there is a double-degenerate vacuum $|\uparrow \rangle$,
$|\downarrow \rangle$, which satisfies
\begin{eqnarray}
c_0 |\downarrow \rangle & =& |\uparrow \rangle \qquad
b_0|\uparrow \rangle = |\downarrow \rangle \,,
\nonumber\\
c_0|\uparrow \rangle & =& b_0|\downarrow \rangle =
c_{n}|\uparrow,\downarrow \rangle  = b_{n}|\uparrow,\downarrow
\rangle=0 \,, \qquad n>0\,.
\label{ghvac}
\end{eqnarray}
On the vacua $|\uparrow \rangle $, $|\downarrow \rangle$ we have
$L_0^{(b,c)}=-1$.  The zero-modes algebra implies that the inner
product on the ghost space is consistently defined~by
\begin{eqnarray}
\langle \uparrow | c_0|\uparrow \rangle & = &\langle \uparrow |
c_0|\uparrow \rangle =1 \,,
\nonumber\\
\langle \uparrow | c_0|\downarrow \rangle & = 0 \,,
\label{ghinp}
\end{eqnarray}
and from this follows the inner product of any excited state.

The BRST charge
\begin{equation}
\label{brstcharge}
Q = \oint dz c(z) \left( T^{M}(z) + \frac12 T^{(b,c)}(z) \right) \,,
\end{equation}
is an hermitean operator which carries ghost number $+1$, and $Q^2=0$
if $c_M=26$. Here $T^M$ is the matter stress tensor.  In this work we
consider the restriction to the space $\cF$ defined by $b_0=L_0=0$,
where $L_0$ is the zero mode of $T=T^M + T^{(b,c)}$.  This chooses the
$|\downarrow \rangle$ vacuum, which will be denoted $|0 \rangle_g$. In
$\cF$ we redefine $N_g$ to be in $\mathbb{Z}$ by assigning ghost
number $0$ to $|0 \rangle_g$ and ignoring the last two terms in the
expansion~(\ref{ng}).  The action of $Q$ is consistent with the
restriction to $\cF$ because $\{ Q,b_0 \} = L_0$.

\subsection{Fermionic strings}\label{section6.2}

For NSR strings, we have, on top of the $(b,c)$ system, the
super-reparametrization ghosts $(\beta,\gamma)$, with conformal
dimensions $(3/2,-1/2)$ and OPEs
\begin{eqnarray}
\gamma(z)\beta(w) &\sim & {1 \over z-w} \,,
\nonumber\\
\beta(z)\gamma(w) &\sim & {-1 \over z-w} \,,
\nonumber\\
\beta(z)\beta(w) &\sim & \gamma(z)\gamma(w) \sim \, 0 \,.
\label{opebg}
\end{eqnarray}
The modes of $(\beta,\gamma)$ satisfy $\beta_r^{\dagger}=-\beta_{-r}$,
$\gamma_r^{\dagger}=\gamma_{-r}$, where $ r \in \mathbb{Z}$  ($r \in
\mathbb{Z}+ 1/2$) for the Ramond (Neveu-Schwarz) sector.  In the
Ramond sector, the zero modes again make the vacuum double-degenerate.
We choose $\beta_0=0$ in $\cF$, and denote the selected vacuum also by
$|0 \rangle_g$.  We have then,
\begin{equation}
\label{vbg}
\beta_{r \geq 0}|0 \rangle_{g} = \gamma_{r>0}|0 \rangle_{g} =0\,.
\end{equation}
On $|0 \rangle_g$, we have $L_0^{(\beta,\gamma)}=3/8$
($L_0^{(\beta,\gamma)}=1/2$) for the Ramond (Neveu-Schwarz) sector.
The $|0 \rangle_g$ ghost number is zero, and the operator $N_g$
in~(\ref{ng})\ is extended to satisfy
\begin{eqnarray}
[N_g,\beta_r]&=& - \beta_r \,,
\nonumber\\
{}[N_g, \gamma_r]&=& + \gamma_r \,.
\label{cargasbg}
\end{eqnarray}
The BRST charge is now
\begin{equation}
\label{brstchfer}
Q = \oint dz \, \left[ c(z) \left( T^{M}(z) + \frac12 T^{g}(z) \right)
+ \gamma(z)\left( G^{M}(z) + \frac12 G^{g}(z) \right) \right] ,
\end{equation}
with $Q^2=0$ if $c_M=10$. Here $G^M$ is the matter supercurrent,
$T^{g}=T^{(b,c)} +T^{(\beta,\gamma)}$ and $G^g$ the supercurrent of
the superconformal combined $(b,c)-(\beta,\gamma)$ systems. The
$L_0=0$ condition is imposed with $T^M + T^g$.  The space $\cF$ is
again defined by $b_0=L_0=0$. In the Ramond sector, also by
$\beta_0=G_0=0$, where $G_0$ is the zero mode of $G= G^M +
G^{g}$. This restriction is consistent because $\{Q,\beta_0 \} = G_0$.

\section{The action of $\cC$ in the ghost sector}\label{section7}

In this appendix we will show that the operator $\cC$, defined in
section~\ref{section2}, interchanges between $b_n$ and $c_n$ (and
$\beta_r$ and $\gamma_r$) modes, thus mapping $\cF_g$ to $\cF_{-g}$.

We can decompose the ghost sector $\cF_{(b,c)}$ into
\begin{equation}
\cF_{(b,c)} = \bigotimes_{n=1}^{\infty} \left\{  |0 \rangle_g, |+,n
\rangle \right\} \otimes \{  |0 \rangle_g, |-,n \rangle \} \,,
\label{ghostdecom}
\end{equation}
where
\begin{equation}
\label{gemasmenos}
|\pm,n \rangle \equiv \frac{ ( b_{-n} \pm c_{-n}) }{\sqrt{2}} |0
 \rangle_g \,,
\end{equation}
satisfy
\begin{eqnarray}
\langle \pm,n | \pm,m \rangle &=&  \pm \delta_{n,m} \,,
\nonumber\\
\langle +,n | -,m \rangle &=&  0 \,,
\label{gein}
\end{eqnarray}
so from $ \cC | \pm,m \rangle= \pm | \pm,m \rangle$ we have
\begin{eqnarray}
\cC c_n \cC & =& b_n \,,
\nonumber\\
\cC b_n \cC & =& c_n \,.
\label{cenghost}
\end{eqnarray}
For fermionic strings, we use the decomposition
\begin{equation}
\cF_{(\beta,\gamma)} = \bigotimes_{r>0} \bigoplus_{N_r=0}^{\infty} \{
|+,N_r \rangle \} \otimes \bigotimes_{r>0} \bigoplus_{N_r=0}^{\infty}
\{ |-,N_r \rangle \} \,,
\label{ghdemm}
\end{equation}
where
\begin{equation}
\label{gemm}
|\pm,N_r \rangle \equiv \frac{ ( \beta_{-r} \pm \gamma_{-r})^{N_r}
 }{\sqrt{2N_r!}} |0 \rangle_g \,,
\end{equation}
satisfy
\begin{eqnarray}
\langle \pm,N_r | \pm,N_s \rangle &=& (\pm 1)^{N_r}
\delta_{r,s}\delta_{N_r,N_s} \,,
\nonumber\\
\langle +,N_r | -,N_s \rangle &=&  0 \,,
\label{geimm}
\end{eqnarray}
so we have
\begin{eqnarray}
\cC \beta_r \cC & =& (-1)^{S(r)} \gamma_r \,,
\nonumber\\
\cC \gamma_r \cC & =& (-1)^{S(r)} \beta_r \,,
\label{cenghmm}
\end{eqnarray}
where $S(r)$ equals $0 \, (1)$ if $r<0 \, (r>0)$.

\section{$\SL(2,\R)/\U(1) \times \U(1)$ character decomposition}\label{section8}

In this appendix we will prove explicitly the character
decompositions~(\ref{cardeco})\ and~(\ref{cardecof}).  Let us start
with the representation $\cD{j}^+$, whose character is
\begin{equation}
\label{carchi}
\chi^+_j(q,z)= { q^{- {j(j-1) \over k-2}} z^j \over
\prod^{\infty}_{n=1}(1-q^n)(1-q^{n-1}z)(1-q^nz^{-1})} \,.
\end{equation}
We will use the technique of infinite partial fractions, which has
been used in~\cite{prod,thorn} in order to prove similar
identities. Define
\begin{equation}
\label{efene}
f_{N}(q,z)= {1 \over {\prod}_{n=1}^{N} \lp 1-q^{n-1}z\rp \lp
1-q^nz^{-1} \rp } \,,
\end{equation}
which can be decomposed into partial fractions as
\begin{equation}
\label{partialefe}
f_{N}(q,z)=\sum^N_{k=1} {a_k(q) \over
( 1 - q^{k-1}z )( 1 - q^kz^{-1})}\,.
\end{equation}
The coefficients $a_k(q)$ are
\begin{eqnarray}
a_k(q) & =& \lim_{z \rightarrow q^k} ( 1 - q^{k-1}z )( 1 - q^kz^{-1} )
f_{N}(q,z) \,,
\nonumber\\
& =& { 1 - q^{2k-1} \over \prod_{n=k}^{N+k-1}(1-q^n)\prod^N_{n=1 n
 \neq k}(1-q^{n-k})} \,,
\nonumber\\
& =& {(1-q^{2k-1}) (-1)^{k-1} q^{{k(k-1) \over 2}} \over
 \prod^{N+k-1}_{n=1}(1-q^n) \prod^{N-k}_{n=1}(1-q^n)} \,,
\label{aka}
\end{eqnarray}
thus for $N \rightarrow \infty$ we have
\begin{eqnarray}
\lim_{N \rightarrow \infty} f_N(q,z) & =& {1 \over
{\prod}_{n=1}^{\infty} ( 1-q^{n-1}z)(1-q^nz^{-1} )} \,,
\nonumber\\
& =& {1 \over \prod^{\infty}_{n=1}(1-q^n)^2 }
\left[ \sum_{k=1}^{\infty} {(1-q^{2k-1})(-1)^{k-1} q^{{k(k-1)
\over 2}} \over (1-q^{k-1}z)(1-q^kz^{-1})} \right] .
\label{efelimite}
\end{eqnarray}
The denominators in the sum of~(\ref{efelimite})\ can be expanded and
resummed as
\begin{eqnarray}
{1 \over (1-q^{k-1}z)(1-q^k z^{-1})} & =& \sum^{\infty}_{m,n=0}
q^{m(k-1)}z^m q^{nk}z^{-n} \,,
\nonumber\\
& =& {1 \over (1-q^{2k-1})} \left[ \sum^{\infty}_{n=0} q^{n(k-1)} z^n
+ \sum^{-1}_{n=-\infty} q^{-nk} z^{n} \right] .
\label{sumageo}
\end{eqnarray}
Thus~(\ref{efelimite})\ can be written as
\begin{eqnarray}
{1 \over {\prod}_{n=1}^{\infty} ( 1-q^{n-1}z)(1-q^nz^{-1} )} & =&
{1 \over \prod^{\infty}_{n=1}(1-q^n)^2 } \times
\nonumber\\&&
\times \biggl[ \sum^{\infty}_{n=0} \sum_{k=1}^{\infty}    z^n
  (-1)^{k-1}  q^{\frac12 (k + 2n)(k-1)}+
\nonumber\\&&
             \hphantom{\times \biggl[}\!
+ \sum_{n=-\infty}^{-1} \sum_{k=1-2n}^{\infty} z^n (-1)^{k-1}
q^{\frac12 (k + 2n)(k-1)} \biggr] \,,\qquad
\label{efelimitedos}
\end{eqnarray}
where in the second term we have changed the index $k$ to $k+2n$ in
order to make the exponents of $q$ be the same in both terms.  Now,
using that for $n \in \mathbb{Z}^-$
\begin{equation}
\label{sumacero}
\sum_{k=1}^{-2n}  (-1)^{k-1} q^{\frac12 (k + 2n)(k-1)} = 0 \,,
\end{equation}
and changing the index $k$ to $s=k-1$ we obtain finally
\begin{equation}
\label{igualdadfinal}
{1 \over {\prod}_{n=1}^{\infty} ( 1-q^{n-1}z)(1-q^nz^{-1} )} = {1
\over \prod^{\infty}_{n=1}(1-q^n)^2} \sum^{\infty}_{n=-\infty}
\sum^{\infty}_{s=0} z^{n}(-1)^s q^{\frac12 s(s+2n+1)} \,,
\end{equation}
and multiplying both sides by $q^{- {j(j-1) \over k-2}} z^j
\prod^{\infty}_{n=1}(1-q^n)^{-1}$, the first line of~(\ref{cardeco})\
follows.

From~(\ref{cardf}), we see that the flowed characters can be obtained
from $\chi^+_j(q,z)$ by replacing $z \rightarrow zq^{-w}$, and
multiplying the resulting expression by $z^{kw/2}q^{-kw^2/4}$.  So we
have,
\begin{eqnarray}
\chi^w_j(q,z)&= & {q^{- {j(j-1) \over k-2}-kw^2/4} \, z^{kw/2} \over
\prod^{\infty}_{n=1}(1-q^n)^3} \sum^{\infty}_{n=-\infty}
\sum^{\infty}_{s=0} (zq^{-w})^{n+j}(-1)^s q^{\frac12 s(s+2n+1)} \,,
\nonumber\\
&= & {q^{- {j(j-1) \over k-2}} z^{j+kw/2} \over
\prod^{\infty}_{n=1}(1-q^n)^3} \sum^{\infty}_{n=-\infty}
\sum^{\infty}_{s=0} q^{{(j+n)^2\over k}-{(j+n+kw/2)^2\over k}} z^n
(-1)^s q^{\frac12 s(s+2n+1)} \,,\qquad
\label{uedt}
\end{eqnarray}
and this identity is the first line of~(\ref{cardecof}).

The decompositions for $\cCo{j,\alpha}$ representations are immediate
and are left as an exercise.

\end{document}